\documentclass[12pt]{iopart}

\usepackage[pdftex]{graphicx}
\setkeys{Gin}{clip=true,draft=false}
\graphicspath{{./pdf/}}
\DeclareGraphicsExtensions{.pdf}
\makeatletter
\def\vec@style{\relax} 
\def\vec#1{\relax\ifmmode\mathchoice
{\mbox{\boldmath$\vec@style\displaystyle#1$}}
{\mbox{\boldmath$\vec@style\textstyle#1$}}
{\mbox{\boldmath$\vec@style\scriptstyle#1$}}
{\mbox{\boldmath$\vec@style\scriptscriptstyle#1$}}\else
\hbox{\boldmath$\vec@style\textstyle#1$}\fi}

\def\mat@style{\sf} 

\def\mat#1{\relax\ifmmode\mathchoice
{\mbox{\boldmath$\mat@style\displaystyle#1$}}
{\mbox{\boldmath$\mat@style\textstyle#1$}}
{\mbox{\boldmath$\mat@style\scriptstyle#1$}}
{\mbox{\boldmath$\mat@style\scriptscriptstyle#1$}}\else
\hbox{\boldmath$\mat@style\textstyle#1$}\fi}
\makeatother
\newcommand{\R}{{\ifmmode\mathbb{R}\else$\mathbb{R}$\fi}}
\newcommand{\C}{{\ifmmode\mathbb{C}\else$\mathbb{C}$\fi}}

\begin{document}
\title{Dynamic Computing Random Access Memory}
\author{F. L. Traversa$^{1,2}$, F. Bonani$^{3}$, Y. V. Pershin$^4$, M. Di Ventra$^2$}
\address{$^1$ Department of Electronic Engineering, Universitat Aut\`{o}noma de Barcelona, Spain}
\address{$^2$ Department of Physics, University of California, San Diego, La Jolla, California 92093-0319, USA}
\address{$^3$ Dipartimento di Elettronica e Telecomunicazioni, Politecnico di Torino, 10129 Torino, Italy}
\address{$^4$ Department of Physics and Astronomy, University of South Carolina, Columbia, South Carolina 29208, USA}
\eads{\mailto{fabio.traversa@polito.it}, \mailto{fabrizio.bonani@polito.it}, \mailto{pershin@physics.sc.edu}, \mailto{diventra@physics.ucsd.edu}}
\begin{abstract}
The present von Neumann computing paradigm involves a significant amount of information transfer between a central processing unit (CPU) and memory, with concomitant limitations in the actual execution speed. However, it has been recently argued that a different form of computation, dubbed {\it memcomputing} [\textit{Nature Physics} \textbf{9}, 200-202 (2013)] and inspired by the operation of our brain, can resolve the intrinsic limitations of present day architectures by allowing for computing {\it and} storing of information on the {\it same} physical platform. Here we show a simple and practical realization of memcomputing that utilizes easy-to-build memcapacitive systems. We name this architecture Dynamic Computing Random Access Memory (DCRAM). We show that DCRAM provides {\it massively-parallel} and {\it polymorphic} digital logic, namely it allows for different logic operations with the same architecture, by varying only the control signals. In addition, by taking into account realistic parameters, its energy expenditures can be as low as a few fJ per operation. DCRAM is fully compatible with CMOS technology, can be realized with current fabrication facilities, and therefore can really serve as an alternative to the present computing technology.
\end{abstract}
\maketitle

\section{Introduction}
\label{intro}

There is currently a surge of interest in alternative computing paradigms \cite{ITRS} that can outperform or outright replace the present von Neumann one \cite{VonNeumann}.
It is clear that such alternatives have to fundamentally depart from the existing one in both their execution speed as well in the way they handle information.
For at least a couple of decades, quantum computing \cite{review_QC,book_QC} (QC) has been considered a promising such alternative, in view of its intrinsic massive parallelism afforded by the superposition principle of quantum mechanics. However, the range of QC applications is limited to a few problems such as integer factorization \cite{Shore} and search \cite{Grover}.

In order to obtain a paradigm shift we then need to look somewhere else but no farther than our own brain. This amazing computing machine is particularly suited for massively-parallel computation. It is polymorphic, in the sense that it can perform different operations depending on the input from the environment, and its storing and computing units -- the neurons and their connections, synapses \cite{synapses} -- are the {\it same} physical object. Such a brain-inspired computing paradigm has been named {\it memcomputing} \cite{DiVentra_NatPhys} and relies on resistors \cite{Chua71,Chua76}, capacitors or inductors with memory (collectively called memelements) \cite{DiVentra_ProcIEEE,DiVentra_Nano} both to store the data and to perform the computation. The features of memelements that make them very attractive from a practical point of view are: {\it i}) they are a natural by-product of the continued miniaturization of electronic devices, and {\it ii}) they can be readily fabricated \cite{Pershin_review,Waser,Sawa,WeiLu} making memcomputing a realistic possibility.

This work reports a memcomputing implementation based on solid-state memcapacitive systems \cite{DiVentra_PRB} (capacitors with memory). While previous memcomputing schemes \cite{memc1,memc2,memc3,memc4,williams,memc5,memc6,memc7} employ intrinsically dissipative memristive devices \cite{Chua71,Chua76} (resistors with memory), we take advantage of very low power dissipation in memcapacitive systems \cite{DiVentra_ProcIEEE} to build a \textit{Dynamic Computing Random Access Memory} (DCRAM) capable of storing information and performing polymorphic logic computation. This new platform allows for massively-parallel logic operations directly in memory thus offering a starting point for a practical solution to the von Neumann bottleneck problem \cite{Backus}. Moreover, we would like to emphasize that our idea is not limited to the specific type of memcapacitive systems used for model calculations reported in this work. For example, ferroelectric capacitors \cite{FECAP} used in FERAM \cite{FERAM} and currently evaluated for new DRAM solutions are also promising candidates for DCRAM.

\begin{figure*}
\begin{center}
\includegraphics[width=0.67\columnwidth]{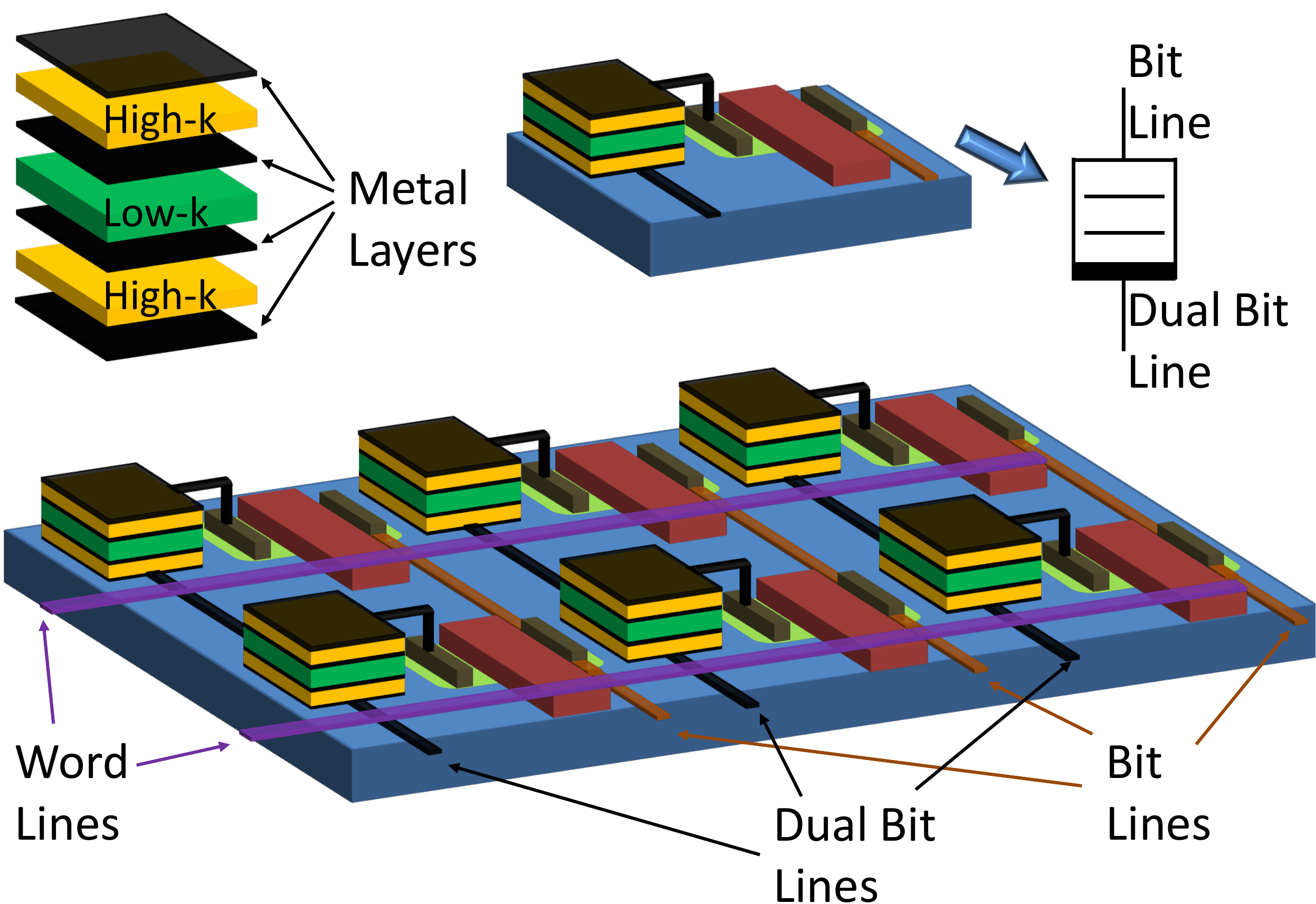}
\end{center}
\caption{\label{dcram}Possible realization of DCRAM. The memory cell (in the top right corner) is a solid-state memcapacitive system composed by three insulating layers separated by metal layers. 
Two-dimensional DCRAM circuit (bottom) is composed by an array of cells having an access element (MOSFET) with a gate controlled by the word line.  In order to perform READ or WRITE operations with a given cell, a positive voltage is applied to its word line, ground to its dual bit line, and suitable voltage pulses to its bit line. For computation purposes, several cells can be coupled through bit and dual bit lines as described in the text.}%
\end{figure*}

While the general topology of DCRAM (Fig.~\ref{dcram}) is similar to that of conventional dynamic random access memory (DRAM), its memory cells are solid-state memcapacitive systems \cite{DiVentra_PRB}. These are multilayer structures composed of insulating layers (three in the particular realization we consider here) alternated by metal layers. The most external insulating layers are made of high-$\kappa$ materials with very high potential barrier so that negligible charge can pass trough them. On the other hand, the intermediate layer is formed out of a low-$\kappa$ material with low potential barrier. This choice allows for non-negligible charge migration between two internal metal layers at appropriate bias conditions. If the middle insulator layer is thin enough, the internal charge current is due to quantum tunnelling \cite{TunnelingBook} and can be easily tuned over a wide range of values \cite{simmons}.

Although no prototype of solid-state memcapacitive systems has been realized yet, we point out that its actual realization oriented to VLSI circuits may not be of a simple planar geometry. In fact, DRAM capacitors are normally of cylindrical shape. Consequently, a possible realization of solid-state memcapacitive systems could consist of three cylindrical capacitors forming an effective solid-state memcapacitive system.

\section{Example of memcapacitor structure and device optimization}

The capacitance $C_d$ of the solid-state memcapacitive system we consider here is defined using the standard relation $q = C_dV_C$, where $q$ is the charge on the capacitor plates (external metal layers) and $V_C$ is the voltage applied to the system. Importantly, $C_d$ is a function of the internal state, namely, it depends on the ratio $Q/q$ where $Q$ is the internal charge (see the top left inset in Fig.~\ref{response} below) \cite{DiVentra_PRB}. Moreover, $C_d$ can diverge and take negative values \cite{DiVentra_PRB} leading to a variety of transient responses.

The internal memory of the memcapacitive system \cite{DiVentra_PRB} arises from the delay of the internal charge response caused by a tunneling barrier of the central insulator layer \cite{DiVentra_PRB}. The tunneling barrier can be lowered by a voltage bias applied to the capacitor plates. In this case, a finite internal current (between the internal metal layers) changing $Q$ is possible. The internal charge $Q$ becomes trapped when the shape of the potential barrier is restored. Therefore, the applied voltage pulses can be used to control the internal charge $Q$, which can be subsequently stored. 

Here we discuss the features of the solid-state memcapacitor as proposed in \cite{DiVentra_PRB}, using realistic values of parameters compatible with the 2012 International Technology Roadmap for Semiconductors (ITRS) specifications \cite{ITRS}. From ITRS 2012, the capacity of DRAM cell is about $20-25$ fF and the equivalent oxide thickness (EOT) is $0.5$ nm for a high-$k$ material of $k=50$. A rapid calculation shows that the area of the metallic layers of an \textit{equivalent} planar capacitor (common geometries for DRAM capacitors are not in general planar, several complex geometries, e.g., cylindrical or pedestal structures, are employed by different manufacturers) has to be of the order of $0.25$ $\mu$m$^{2}$, so we use this value in our simulations. Moreover, the physical thickness of the insulator, from the EOT and $k=50$, ranges between $6-10$ nm. Using these data, we consider the memcapacitor structure sketched in Fig.~\ref{dcram}. The thickness of the two high-$k$ insulators is supposed to be $6$ nm and we assume they are made of standard modern high-k material (e.g. TiO$_{2}$) with $k=50$. Finally in our simulations we consider transmission lines with common values for DRAM fabrication, i.e., $R=1.5$ k$\Omega$ mm$^{-1}$ and $C=0.2$ pF mm$^{-1}$ for a length of $1$ mm.

Physical parameters (thickness and $k$ value) of the low-$k$ layer require a more careful consideration since the lifetime of $Q$ strongly depends on these two parameters.
Let us then focus on the \textit{storage mode}, namely, the situation that follows a WRITE operation (application of 1 ns voltage pulse of certain amplitude).
In order to model the least favorable conditions such as a strong external leakage current (due to imperfect switches and other processes), we assume $V_C=0$ irrespective of the written bit. This choice is different from that in common DRAM where, in the storage mode,  $V_C>0$ if the stored bit is 1 and $V_C=0$ if it is 0. Our main goal here is to evaluate the possibility of information storage on long time scales compared to typical DRAM decay times using, however, DRAM-like chip structure.

Let us consider a physical model of solid-state memcapacitive system with a barrier height of 0.2 eV for the low-$k$ material and infinite barrier for the high-$k$ one. The equations governing the time variation of $Q$ and $q$ can be written as \cite{DiVentra_PRB}
\begin{eqnarray}
V_{C} &  =\frac{Q}{C_{2}}+\frac{q}{C_{0}}\\
\frac{dQ}{dt} &  =-I(Q+q) \label{dQdt}
\end{eqnarray}
where $I$ is the tunnel current through the low-$k$ material, $C_{0}$ is the (constant) capacitance of the total memcapacitive system (with respect to $q$ only) and $C_{2}$ is the capacitance of the internal capacitor composed by the low-$k$ material and internal metal layers. If the barrier is sufficiently thin then the current can be approximated by the Simmons formula \cite{simmons}.

\begin{figure*}
\begin{center}
\hfill
\includegraphics[width=0.84\columnwidth]{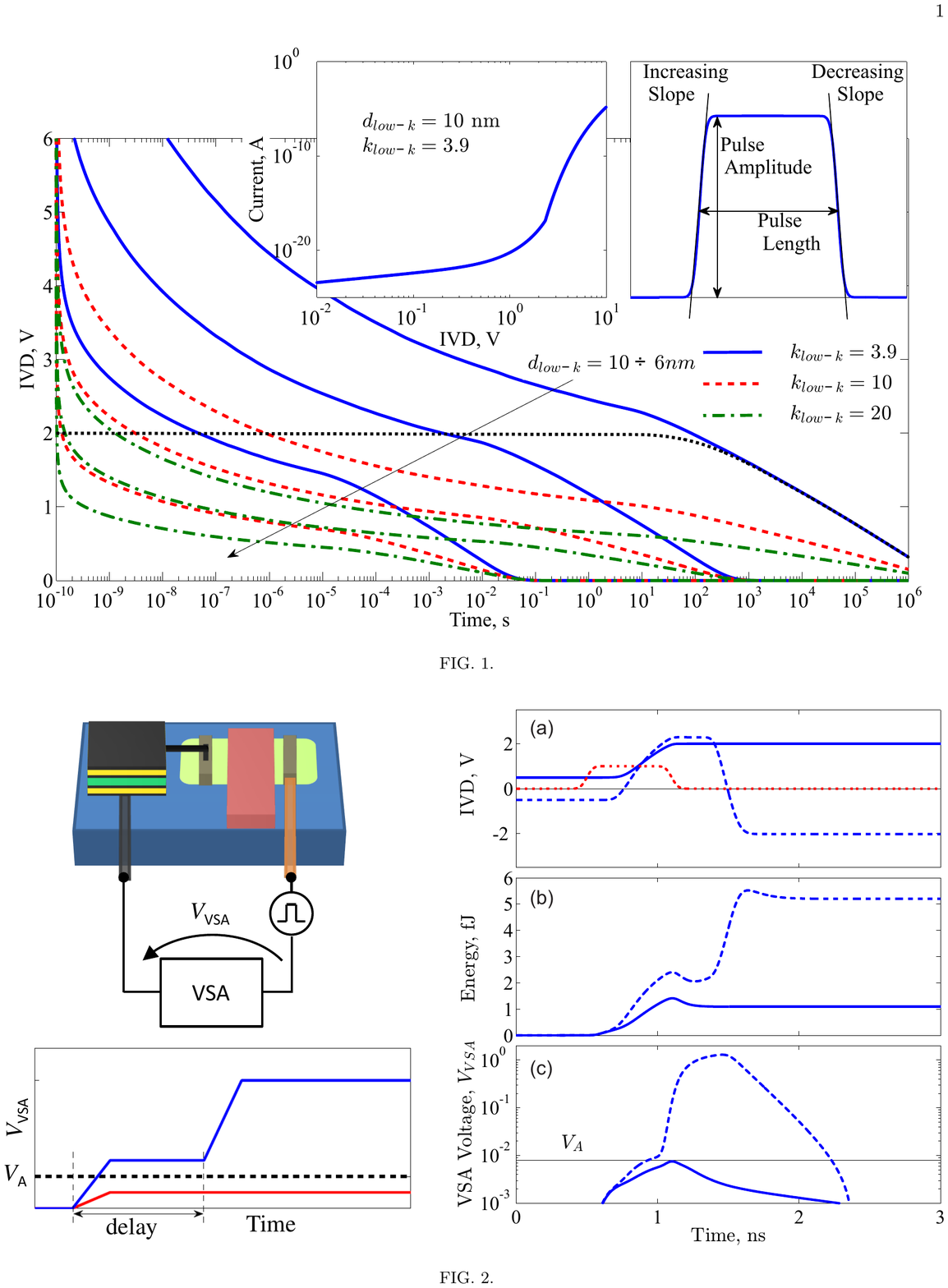}
\end{center}
\caption{\label{optimization}Decay of the internal voltage difference (IVD) $Q/C_2$ for different thicknesses and dielectric constants of the middle insulator.
This plot shows envelope curves of the internal charge decay that can be used to track the long-time behavior for any initial value of $Q$. As an example, the dotted black curve
represents the decay of IVD $Q/C_2$ for the initial condition $Q/C_2=2$V at $k_{low-k}=3.9$ and $d_{low-k}=10$ nm. Note, that this curve converges with the corresponding envelope at longer times. The top left inset reports the current $I((1-C_{0}/C_{2})Q)$ versus $Q/C_2$. Top right inset presents the shape of the voltage pulse with $10$ V/ns rising and falling edges.}
\end{figure*}

Taking into account that $C_{0}<C_{2}$ and $I$ is monotonous with a unique $0$ at $Q+q=0$, there is unique steady-state solution $Q=q=0$ at $V_{C}=0$. The top inset in Fig.~\ref{optimization} shows that the current $I(Q+q)$ is very small at smaller values of $Q+q$  suggesting the possibility of quite low charge relaxation rate at nonzero $Q$.
At $V_{C}=0$, Eq. (\ref{dQdt}) can be rewritten as
\begin{equation}
\frac{dQ}{dt}=-I((1-C_{0}/C_{2})Q).
\end{equation}
This equation describes the decay of the internal charge $Q$ in the storage mode. Fig.~\ref{optimization} shows the decay of $Q$ for several values of $k$ and layer thicknesses. It is worth noticing that at certain values of parameters, such as the thickness of 10 nm and $k=3.9$, the information is stored for a long time. In fact, after $10^{6}$ s (about $11.5$ days) a reasonable amount of charge still remains in the memcapacitive system. Thus, modifying the parameters of the memory cell (the layer thickness, dielectric constant or even the barrier height) one can select an appropriate lifetime of the internal charge $Q$.

\section{WRITE and READ operations.}


In our scheme, the binary information is encoded in the internal charge $Q$ of the memcapacitive system. It is assumed that $Q\geq Q_r$ corresponds to logic $1$, $Q\leq-Q_r$ corresponds to logic $0$, and the logic value is not defined when $-Q_r<Q<Q_r$.  The threshold $Q_r$ is introduced to reliably distinguish logic values, and as such is defined according to the  sensitivity of the voltage sense amplifiers (VSA) that we exploit to allow for the bit value detection.

When a voltage pulse is applied to a memory cell, its current response strongly depends on its internal charge $Q$. We thus use this current response to read the information stored in the memory cell: The common solution (widely used in consumer electronics including standard DRAM technology) employs VSAs.

\begin{figure*}
\begin{center}
\hfill
\includegraphics[width=0.84\columnwidth]{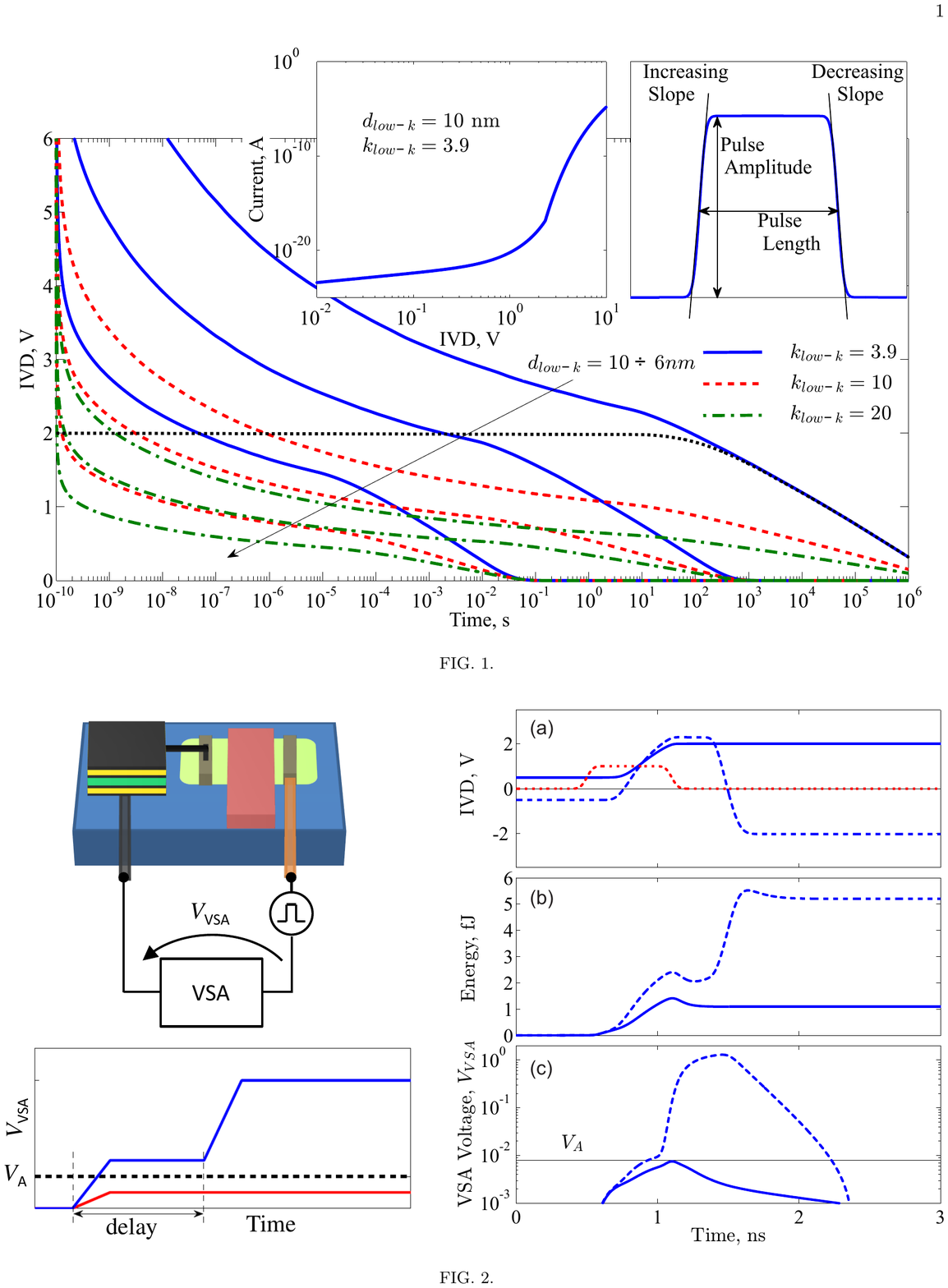}
\end{center}
\caption{\label{VSA}Configuration and simulation of READ-REFRESH process. The circuit configuration for this process is presented on the left. It consists of a memory cell connected to a pulse generator and VSA. The bottom left plot shows the ideal VSA response when its input signal is below (red line) and above (blue line) its threshold. The simulation of a READ-REFRESH process for the initial condition of a partially decayed bit ($Q_{i}/C_{2}=\pm0.5$ V) are given on the right. Here, the circuit is driven by $0.5$ ns length, $1$ V amplitude voltage pulse  during the delay time of VSA. We report (a) the time variation of the normalized charges $Q/C_{2}$ (solid and dashed blue lines) and the voltage pulse (dotted red line), (b) the dissipated energy, and (c) VSA output.}
\end{figure*}

As depicted in Fig.~\ref{VSA}, the VSA is connected to the memory cell in series with a voltage pulse generator. The ideal characteristics of the VSA are presented in the left bottom inset of Fig.~\ref{VSA}. It is important to know that VSA amplifies the response voltage $V_{VSA}$ if $V_{VSA}>V_A$, where $V_A$ is a certain threshold voltage. Generally, the delayed response of VSAs is associated to the internal capacitances of the Metal-Oxide-Semiconductor (MOS) structures they are made of. During the delay time, the voltage pulse generator induces the response voltage $V_{VSA}$. Being amplified, $V_{VSA}$ provides the value stored in the memory cell.

\begin{figure}
\begin{center}
\hfill
\includegraphics[width=0.84\columnwidth]{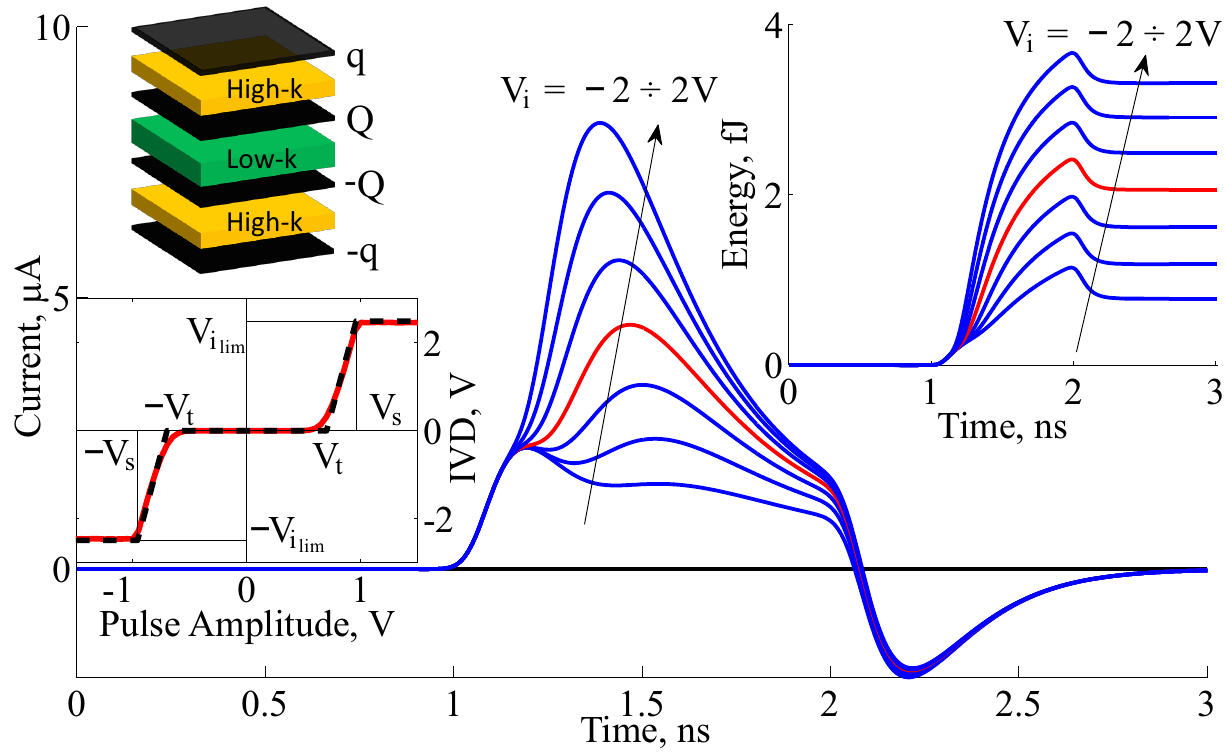}
\end{center}
\caption{\label{response}Single cell response to a voltage pulse under READ/WRITE conditions as described in Fig. \ref{dcram}. In our simulations, the bit and dual bit lines are modeled as transmission lines with typical parameters for DRAM $R=1.5$ k$\Omega$mm$^{-1}$ and $C=0.2$ pFmm$^{-1}$ assuming $1$ mm line length. The voltage pulse is a smooth square pulse of $1$ V amplitude and $1$ ns width starting at $t=1$ ns. The main graph is the current response measured at the end of the bit line for several initial values of the internal charge $Q$. The red line refers to $Q=0$ initial condition. To quantify $Q$, an effective internal voltage difference (IVD) is defined as $V_i=Q/C_2$ with $C_2$ the geometrical capacitance of the intermediate layer, $C_2=A\varepsilon_0k_{low-k}/d_{low-k}$, where $A$ is the surface area, $\varepsilon_0$ is the vacuum permittivity, $k_{low-k}$ is the relative permittivity of the central layer, and $d_{low-k}$ is its thickness. The top right inset shows the cell's dissipated energy. Bottom left inset: the effective internal voltage difference as a function of voltage pulse amplitude in $1$ s after the voltage pulse application. }
\end{figure}

The WRITE, READ and logic operations with memcapacitive memory cells are performed with the help of control circuitry that provides appropriate signals. In order to make the
discussion even more realistic, the parameters we have used throughout the simulations belong to the ITRS 2012
standards \cite{ITRS}. Simulations have been carried out using the general purpose in-house NOSTOS (NOnlinear circuit and SysTem Orbit Stability) simulator developed by one of the authors (FLT) initially for studying circuit stability \cite{TraversaIJCTA,Cappelluti2013}, and recently extended to analyze circuits including memory elements \cite{traversa13}. Let us consider the WRITE operation first. For this purpose, we employ the circuit configuration shown in the top right corner of Fig.~\ref{dcram} where the dual bit line (DBL) is grounded and the voltage pulse is applied to the bit line (BL). As it is mentioned above, applied voltage pulse lowers the potential barrier between the internal metal layers allowing for an internal charge redistribution.

An important observation that one can make at this point is that the WRITE process is of the threshold type. Indeed, one can define a threshold voltage $V_{t}$ such that there is no significant charge transfer between the internal plates at applied voltage amplitudes below $V_{t}$ (see the bottom left inset in Fig.~\ref{response}). On the contrary, at pulse amplitudes exceeding $V_{t}$ a considerable amount of charge can tunnel between the internal layers. In our device structure, $V_{t}$ is about $0.5$ V, which is much larger than the perturbations usually induced by MOS transistor leakage currents. Moreover, the existence of $V_{t}$ results also in an internal charge saturation shown in the bottom left inset of Fig.~\ref{response}.


Next, let us consider the READ operation in DCRAM. Similarly to DRAM, the READ process is destructive (see the top right plot of Fig.~\ref{VSA}: when the voltage pulse acts, the information inside the memory cell is destroyed since the final state inside the memory cell is 1) and thus needs to be followed by a REFRESH operation. In order to have a better understanding, we consider the current response shown in Fig.~\ref{response}.  One can notice significant variations in the cell response depending on the initial value of $Q$. These differences are used to measure the logic value stored in the cell with VSAs similarly to DRAM technology. However, VSA amplifies a voltage difference above or below a certain voltage threshold. To meet the VSA modus operandi, the current response can be transformed into the voltage response connecting the bit and dual bit lines to VSA input terminals. As the voltage pulse used in READ changes the internal charge $Q$, a suitable REFRESH operation is applied after the READ.

\begin{figure*}
\begin{center}
\hfill
\includegraphics[width=0.84\columnwidth]{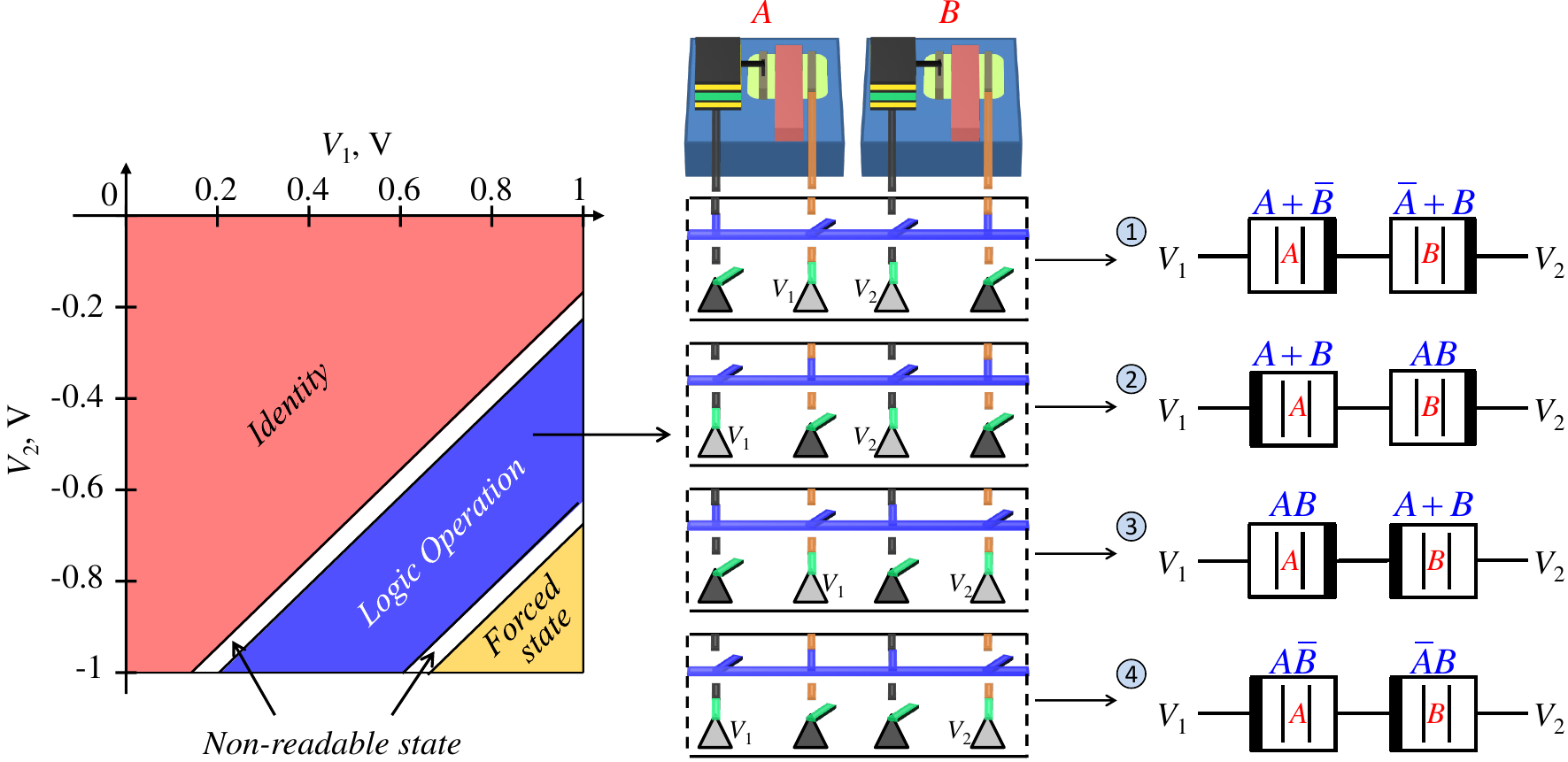}
\end{center}
\caption{\label{computation} Map of logic operations. Two memory cells can be connected in four different ways giving rise to four logic operations. The symbols $+$ and $\bar{\null}$ are the OR and NOT operation respectively, while the AND operation is the implicit multiplication. Here, $V_1$ and $V_2$ are amplitudes of voltage pulses applied to the external connections of the coupled memory cells. Depending on these amplitudes, there are several regions in the logic map. Amplitudes belonging to the \textit{identity} region do not change initial values in memory cells. Amplitudes belonging to the \textit{logic operation} region perform computation as in the scheme to the right. Amplitudes belonging to the \textit{forced state} region change the initial values to 1 or 0 depending on device coupling order and polarity. Amplitudes belonging to the \textit{non-readable state} region produce an intermediate (non readable) internal states with $-Q_r\le Q\le Q_r$.}
\end{figure*}

In summary, the sequence consists of two steps. First, a voltage pulse (in our simulations, of $0.5$ ns length and $1$ V amplitude) is applied by the generator. It produces a voltage response that is considered as input for VSA during its \lq\lq{}delay state\rq\rq{}. Subsequently, if $V_{VSA}>V_A$ the VSA amplifies the voltage $V_{VSA}$ and 0 is written, on the contrary, if $V_{VSA}<V_A$ the VSA does not act and 1 is written. Fig.~\ref{VSA} reports simulations of the READ-REFRESH process considering an extreme case of a partially decayed bit showing all the features mentioned above. Moreover, we would like to emphasize that the dissipated energy has a significant dependence on the value of bit (0 or 1). This is due to an asymmetry in VSA response. In fact, when $V_{VSA}>V_A$ (VSA is activated) the dissipated energy is about $5$ fJ. In the opposite case (initial value is $1$) this energy is about 1 fJ.


The top right inset of Fig.~\ref{response} shows the dissipated energy when a pulse of 1 ns length and 1 V amplitude is applied. In fact, this calculation gives a reference for the order of magnitude of the dissipated energy for all DCRAM operations (WRITE, READ, COMPUTATION) because of close operating conditions. It is worth noticing that this energy is of the order of few fJ, comparable to the best cases of extremely low-energy storage and computation \cite{Zhirnov}, and  computation only with CMOS architectures \cite{Zhirnov}. Importantly, the information is stored directly in DCRAM saving the power usually needed to transfer it to/from the CPU.

\section{Polymorphic computation.}


\begin{figure}
\begin{center}
\hfill
\includegraphics[width=0.84\columnwidth]{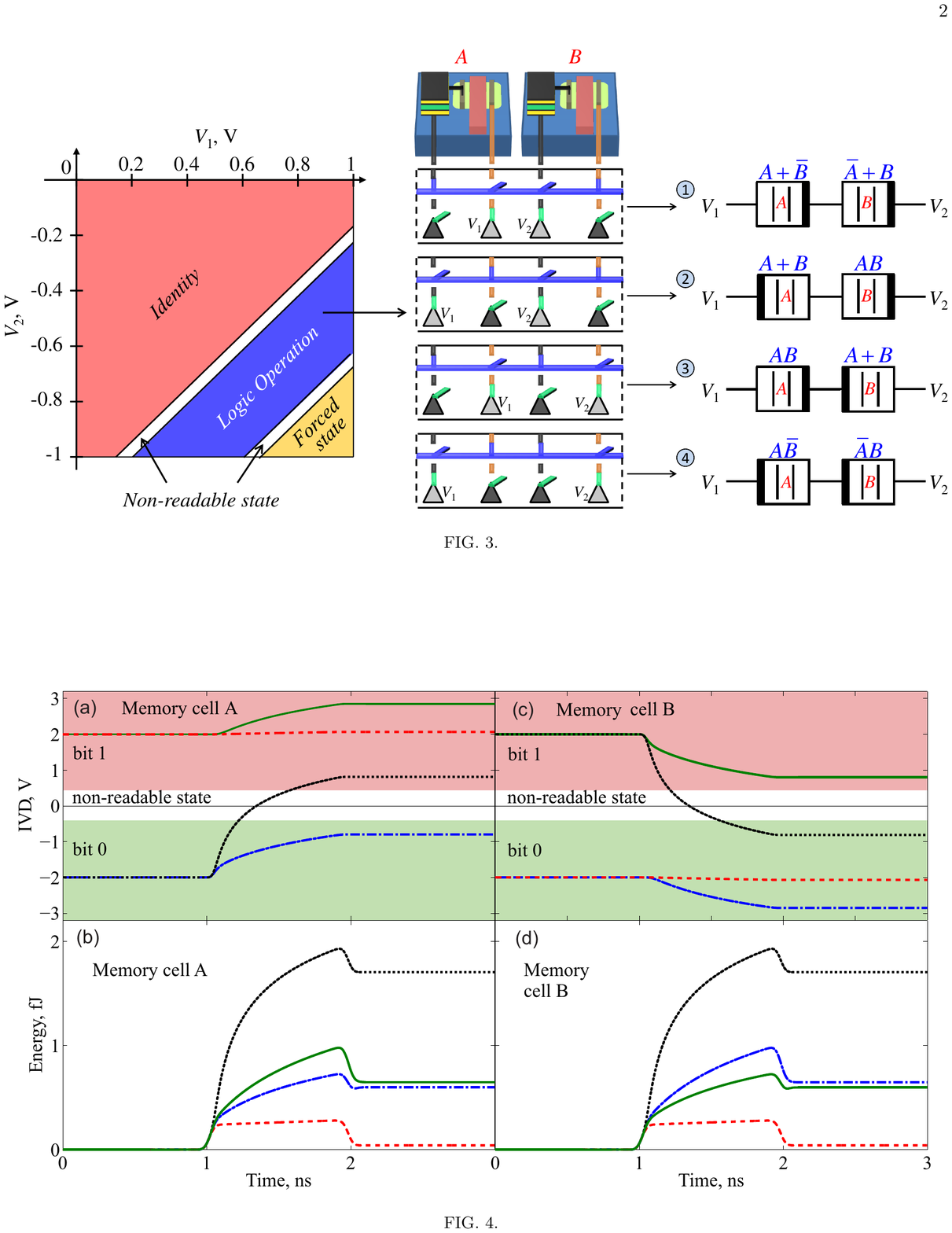}
\end{center}
\caption{\label{computation1}Time variation of IVD and dissipated energy for the second logic gate of figure~\ref{computation}. The voltage pulse amplitudes are $V_1=0.73V$ V and $V_2=-0.73$ V, and the pulse length is 1 ns. The evolution of IVD for both memory cells at different initial conditions are shown by different line styles in (a) and (c). The dissipated energy is plotted in (b) and (d).}
\end{figure}

Let us consider the simplest realization of logic gates when 2 memory cells are used to store the input and (after the computation) the output values. For computation purposes, these memory cells are coupled as shown in Fig.~\ref{computation} using appropriate switches at the end of the BL and DBL. As shown in Fig.~\ref{computation}, the dynamics of the internal charges $Q$ of two coupled cells subjected to a couple of synchronized voltage pulses depends on the initial combination of internal charges of these  cells. In this way, the final values of the internal charges can be considered as a result of a logic operation over initial values stored at $t=0$ in these cells (see fig.~\ref{computation}). As a specific example, let us consider configuration 2 from Fig.~\ref{computation} assuming that $-0.73$ V and $0.73$ V amplitude voltage pulses are applied to the memory cells. Fig.~\ref{computation1} demonstrates the evolution of $Q$ for both cells.  Notice that the final values of $Q$ in cells A and B realize OR and AND gates, respectively. The dissipated energy (bottom plots in Fig.~\ref{computation1}) is quite low: it is less than $2$ fJ in the worst case scenario, and, in the case of ($1$, $0$) initial configuration, it is much lower. However, it is worth noticing that, after computation (see Fig.~\ref{computation1}), the bits stored in the cells are only partially written: the computation process must be completed by a REFRESH process, thus increasing the total required energy per operation by a few fJ, depending on the actual realization of VSA.

Considering possible connections and device polarities one can find that two coupled cells can be used to form a (redundant) basis for a {\it complete set of logic operations}. In fact, it is known \cite{russel, williams} that combining only AND and NOT or OR and NOT functions, any logic expression can be evaluated. In our case, with two coupled memory cells we can in fact perform 6 different two-bit operations, depending both on how the cells are coupled, and on the amplitudes of the applied voltage pulses. Therefore, these two coupled memory cells form universal logic gates as it is exhaustively proved below. 

The universal gate offered by the DCRAM architecture is not its only advantage. DCRAM is capable of intrinsically {\it parallel} computation. In fact, after only one computation step,
we find a different output on each memcapacitive system: this means two operations at the same time. As shown later, by connecting three memory cells and varying the pulse amplitudes and the connection topology, we can perform even more complex operations in one step, and obtain three different outputs written into each memory cell. More importantly, one can perform simultaneous operations over multiple groups of two or three coupled cells. We also point out that by using only one of the possible connection topologies of three memory cells, we obtain another universal gate for two-bit computation with {\it fixed} connection topology representing a possible solution to avoid the supplemental circuitry needed for dynamic connections.

\subsection{Two-bit and Three-bit operations with dynamic connections}

\begin{figure}
\begin{center}
\hfill
\includegraphics[width=0.84\columnwidth]{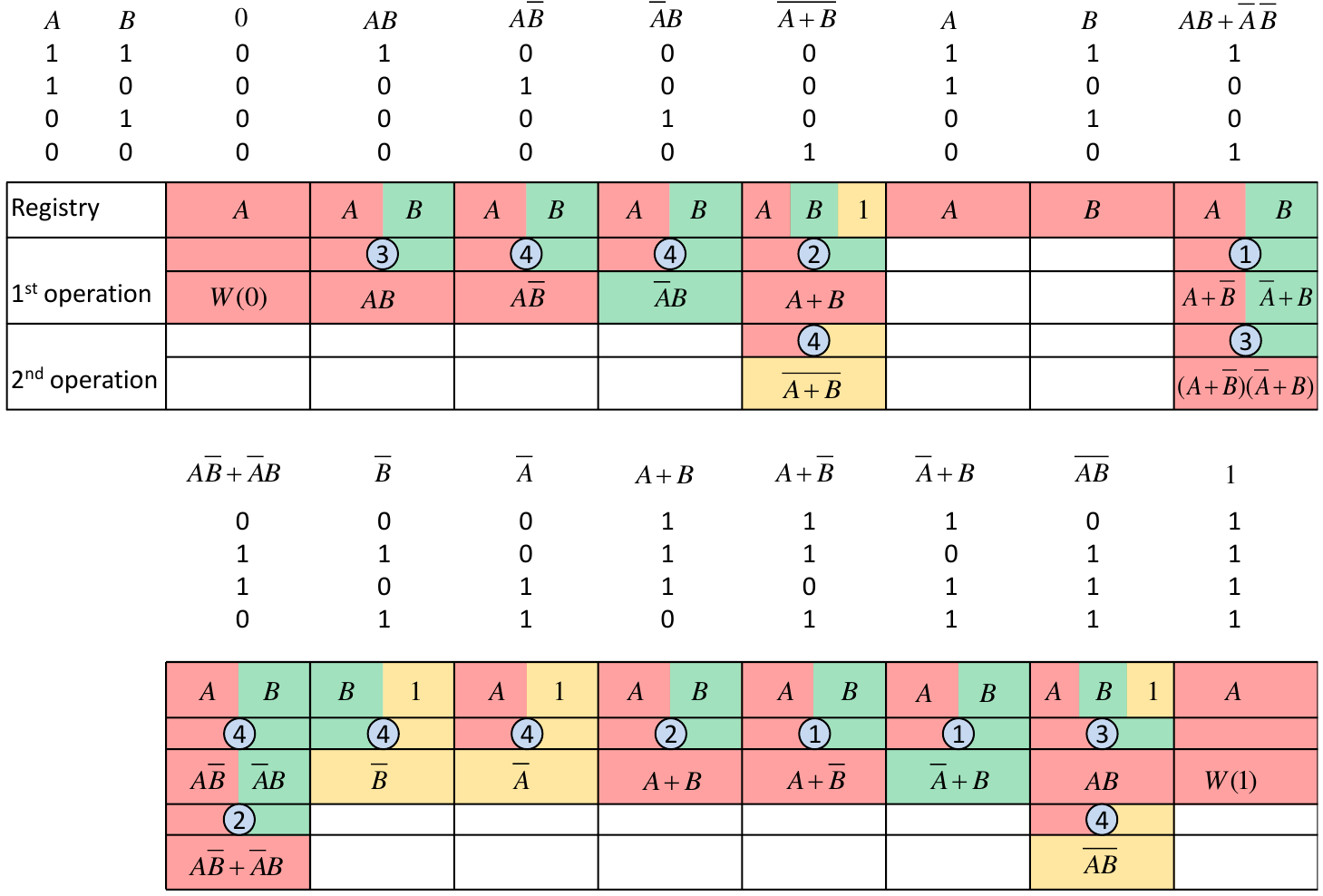}
\end{center}
\caption{\label{logic}Two-bit logic functions. The additional bit set to $1$ is used for negation. Circled numbers refer to logic operation of figure \ref{computation}. Colors denote the  memory cell involved in operation and the cell storing the output. W(1) and W(0) stand for the operation WRITE 1 and 0,  respectively.}%
\end{figure}

Parallel logic operations performed by DCRAMs, summarized in figure~\ref{computation}, can be used to define logic gates forming a (redundant) complete basis for any boolean logic function. In order to prove this claim, figure~\ref{logic} shows how to perform all possible two-bit logic operations using DCRAM gates. We notice that in the worst case scenario, a three-bit registry (three cells) is needed (the third bit, initially set to $1$, is used to perform negation), and a two-level operation is required. Compared with CMOS NAND logic or NOR logic, DCRAM logic circuits require less components. In fact the commonly used CMOS  NAND or NOR logic gates require up to $5$-level operation scheme, and up to $20$ transistors to perform the same set of two-bit functions.

\begin{figure}
\begin{center}
\hfill
\includegraphics[width=0.84\columnwidth]{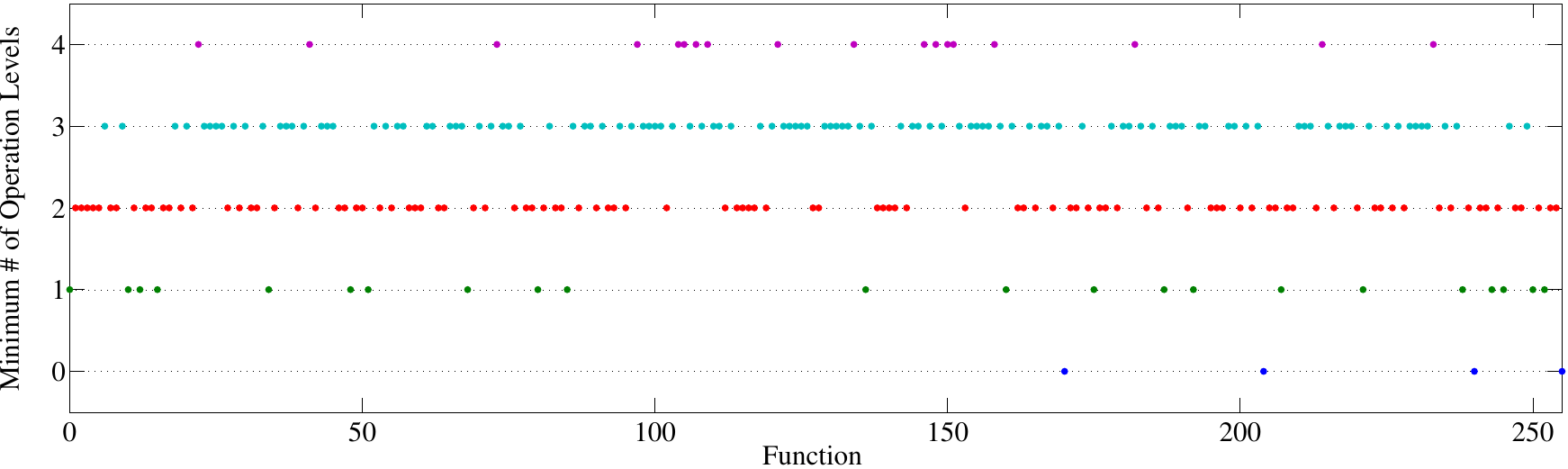}
\end{center}\caption{\label{3bit_comp}Number of operation levels for any three-bit boolean function. There are $2^8$ possible boolean functions involving three bits, so in the $x-$axis each function is coded using the equivalent decimal number.}
\end{figure}

\begin{figure}
\begin{center}
\hfill
\includegraphics[width=0.84\columnwidth]{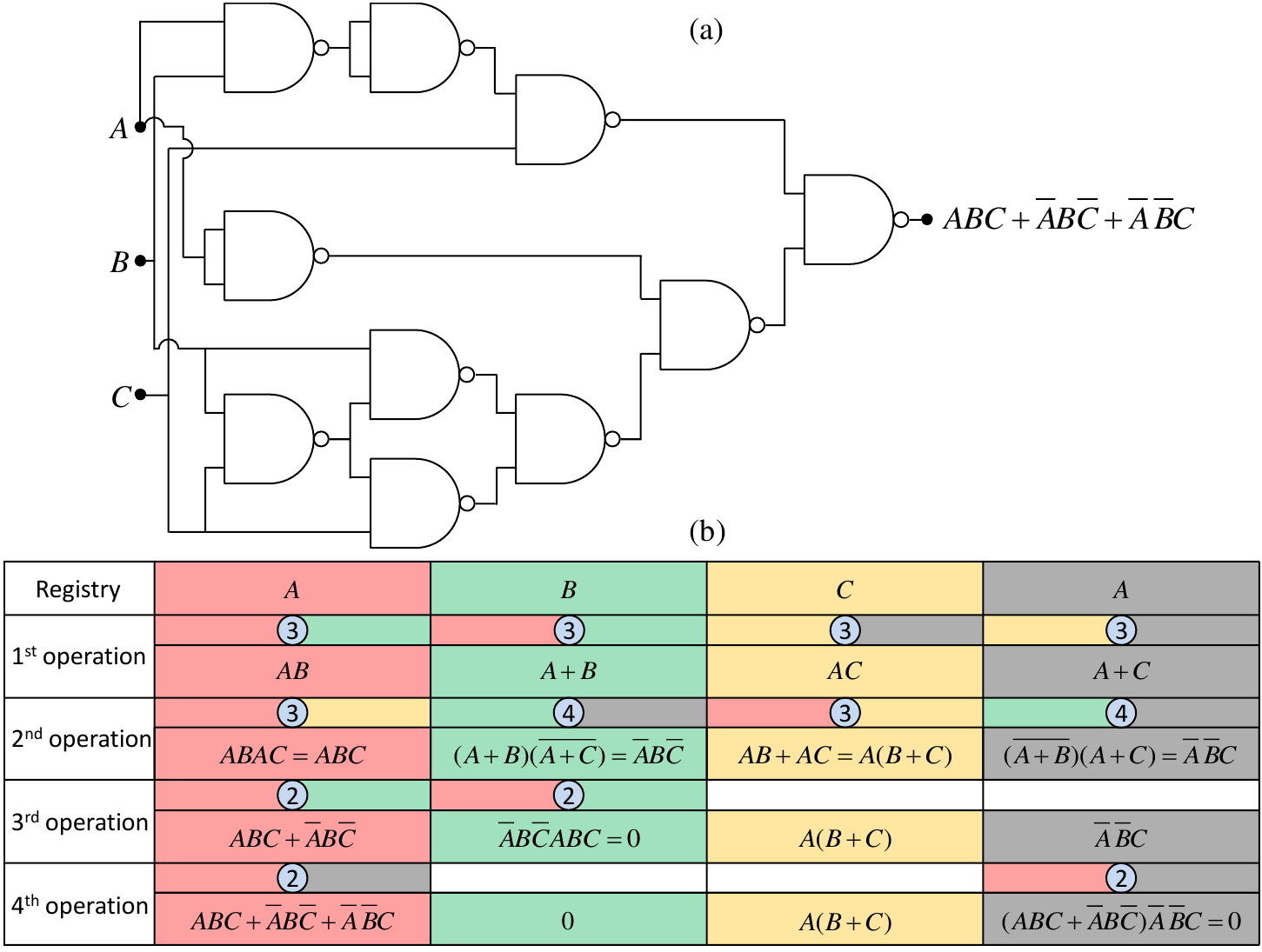}
\end{center}
\caption{\label{logic_3bit}(a) CMOS-NAND logic circuit for the three-bit operation $ABC+\bar{A}B\bar{C}+\bar{A}\bar{B}C$. (b) DCRAM $4$--level scheme for the same logic function.}
\end{figure}

Using the same scheme, we can perform any $n$--bit operation exploiting $2$--bit gates only. Here, we make some considerations on three-bit operations, for which a complete treatment is possible. Using a $5$--bit registry made of the $A$, $B$ and $C$ inputs and two additional bits, one set to $1$ employed for negation and the other equal to one of the three inputs $A$, $B$ or $C$ (depending on the desired logic function), any three-bit logical operation can be performed using at most a $4$--level operation scheme (Fig.~\ref{3bit_comp}). In figure \ref{logic_3bit}--(b) an example of $4-$level three-bit operation is shown. In this case, the registry is composed by the three inputs ($A$, $B$ and $C$) and only one additional bit (in this case $A$), because no bit for negation is required. In figure \ref{logic_3bit}--(a), the comparison with a two-input NAND logic (possibly using programmable digital circuits) is reported. It is worth noticing that, using CMOS NAND logic, the same operation is performed within a $5-$level operation scheme using $10$ NAND gates, i.e., $40$ transistors, thus proving that the complexity of the CMOS circuit is much higher than for our DCRAM implementation.

\subsection{Fixed connection two-bit operations}

\begin{figure}
\begin{center}
\hfill
\includegraphics[width=0.84\columnwidth]{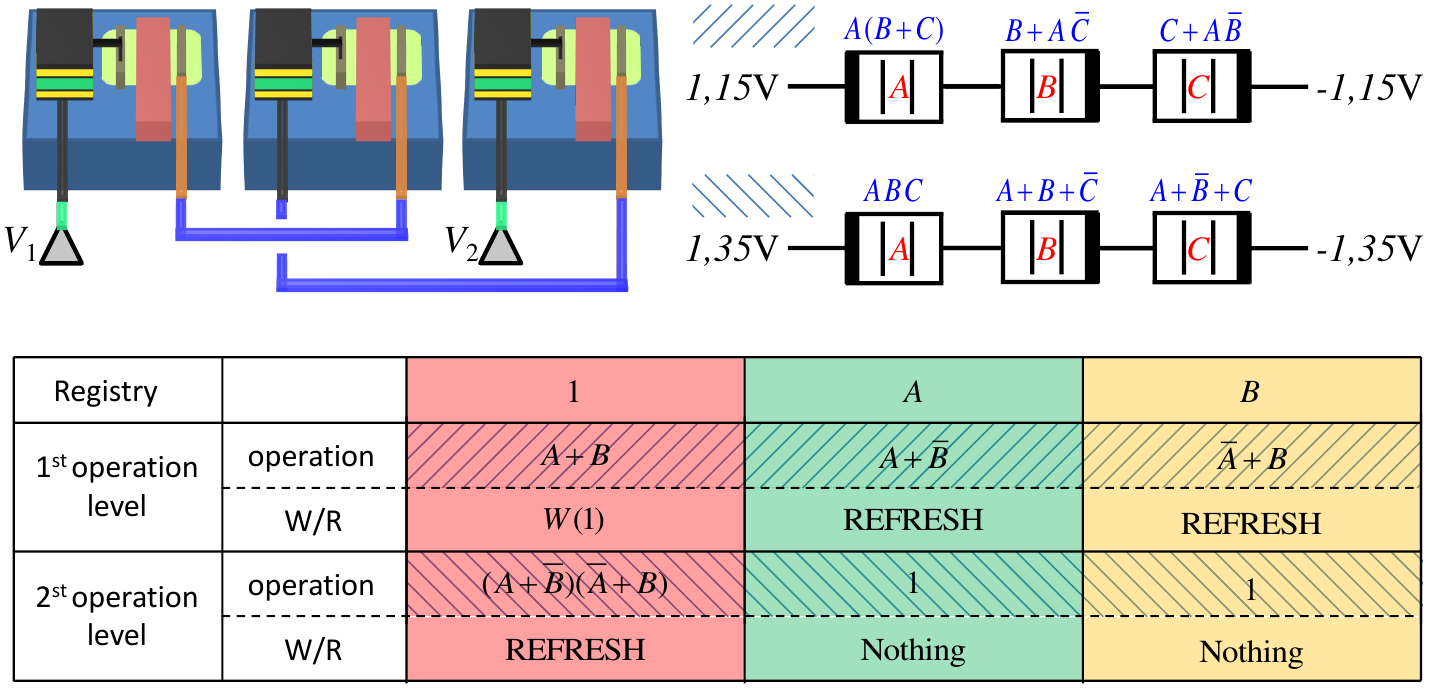}
\end{center}
\caption{\label{2bit_fix}Computation of the logic function $AB+\bar A \bar B$ using three connected memory cells. The topology of the connections is represented in the top left of the figure. The two gates obtained varying the pulse amplitude are sketched on the top right of the figure and indicated by the two different textures on the left of the gates.}
\end{figure}

\begin{figure}%
\begin{center}
\hfill
\includegraphics[width=0.84\columnwidth]{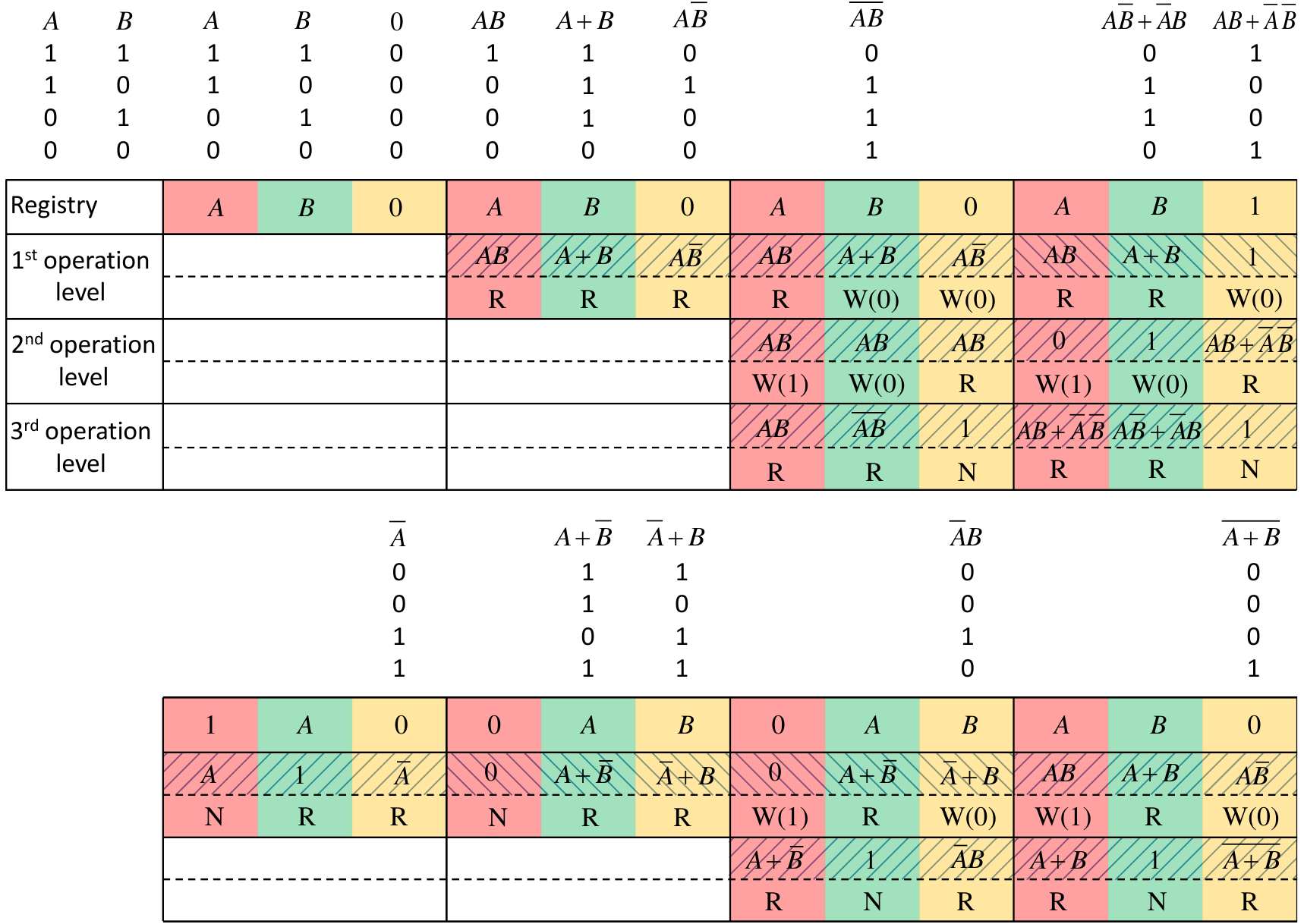}
\end{center}
\caption{\label{2bit_fix_tot}Two-bit logic functions. The configuration of the connections for the three-memory cell polymorphic gate is the same as that in fig. \ref{2bit_fix}. The textures indicate the gate kind as in fig. \ref{2bit_fix} (depending on the pulse amplitudes). W(1) and W(0) stand for the operation WRITE 1 and 0, respectively, and R = REFRESH. The functions $\bar B$ and 1 are not reported for sake of conciseness because they can be simply obtained as in the fifth column for $\bar A$ and in the first column for 0, respectively.}%
\end{figure}

\begin{figure*}
\begin{center}
\hfill
\includegraphics[width=0.84\columnwidth]{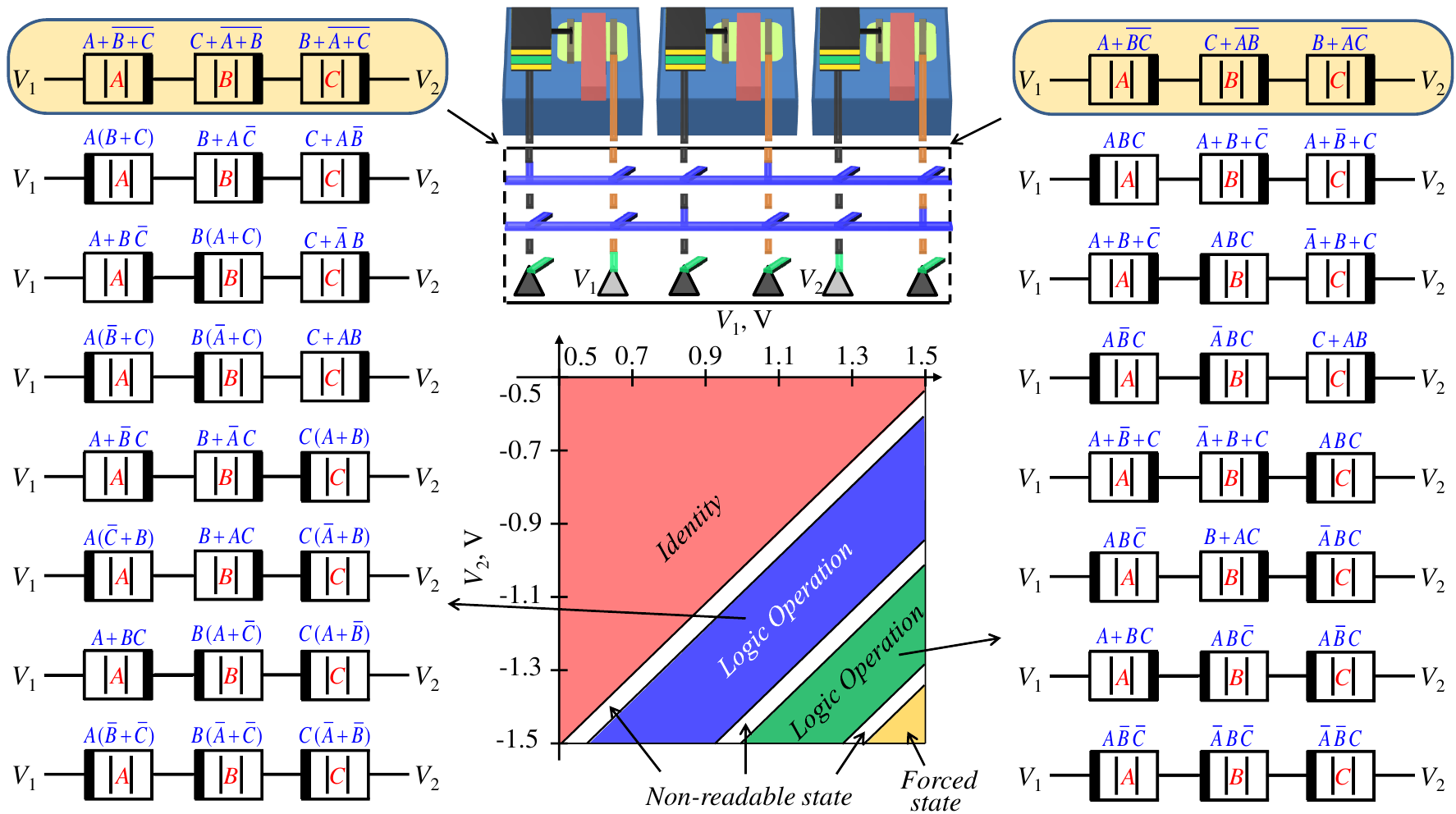}
\end{center}
\caption{\label{3mem}Logic gates with three coupled cells. In the center, we show a map of operations as a function of amplitudes of pulses applied to the external connections of the coupled memory cells. Depending on these amplitudes, there are several regions in the logic map. Amplitudes belonging to the \textit{identity} region do not change initial values. Amplitudes belonging to the \textit{logic operation}  region realize logic functions presented in schemes to the right and left. Amplitudes belonging to the \textit{forced state} region change the initial values to 1 or 0 depending on device coupling order and polarity. Amplitudes belonging to the \textit{non-readable state} region produce an intermediate (non readable) internal states with $-Q_r\le Q\le Q_r$. The symbols $+$ and $\bar{\null}$ are the OR and NOT operations, respectively, the implicit multiplication is the AND.}
\end{figure*}

Finally, we consider the three-bit gate presented in fig.~\ref{2bit_fix}. We assume a configuration with {\it fixed} connections (while computation is performed). As shown in figure fig.~\ref{2bit_fix}, varying the pulse amplitudes applied to the cells we can obtain two different logic outputs for each memory cell. We define these as the logic outputs of the first and second kind. Moreover, at each computation step the REFRESH and WRITE processes are performed to prepare the cells for the next computation step. The bits $1$, $A$ and $B$ are initially written in the three memory cells (registry). Then, we apply the synchronized voltage pulses $V_1$ and $V_2$ with amplitude $1.15$ V and $-1.15$ V, respectively, to obtain the gate of the first kind. The first-level operation is completed by the REFRESH of the second and third memory cells and by writing $1$ in the first one. Then, the second-level operation implements the gate of the second kind, and the boolean function $AB+\bar A \bar B$ is obtained.

Using the processes described above, we can set up a universal gate capable to perform any two-bit logic operation without changing the topology of the circuit. For example Fig.~\ref{2bit_fix_tot} shows how to obtain all possible 2-bit logic functions using the 3-bit fixed polymorphic gate of fig.~\ref{2bit_fix}. Finally, fig.~\ref{3mem} reports the variety of 3-bit polymorphic logic gates that can be implemented using  three coupled memory cells. In this case it is evident the separation into two regions of applied voltage amplitudes providing polymorphism without changing the connection topology.

\section{Conclusions}

In conclusion we have introduced a simple, practical, and easy-to-build memcomputing architecture that processes and stores information on the same physical platform using two-terminal passive devices (memcapacitive systems). Being low-power, polymorphic and intrinsically massively-parallel, DCRAM can significantly improve computing capabilities of the present day von Neumann architecture. This is performed by transferring a significant amount of data processing directly into the memory, where the data is stored. Although it is still an open question which specific algorithms will mostly benefit from such an approach, we expect that our scheme will be extremely useful in scientific calculations, image and video processing and similar tasks.

In order to make a specific estimation of computation speed-up using our approach, let us compare a performance of a traditional personal computer equipped with typical DRAM chips with this of a DCRAM-based computer. For example, consider a 4 GB memory system, with two 2 GB ranks, each consisting of eight 256 MBx8, 4-bank devices \cite{Udipi}. Moreover, each of the 4 banks in a 256 MB device is split into 8 arrays of 8 MB each. If there are 65,536 rows of 1024 columns of bits in each array, a row access provides a selection of 1024 bits per array, giving a total of 65,536 bits across 8 chips of 8 arrays each. This is the number of bits that can be involved  simultaneously in a {\it single} parallel calculation using DCRAM, which lasts for about 20 ns (accounting for a 4-level computation) as discussed above (here we assume that all 65,536 bits are grouped into small few-bits circuits at each calculation step). On the other hand, a standard CPU processes 64 bits per each clock cycle. Accounting for the memory access time of 10 ns \cite{Patterson}, we can conclude from this simple example that a DCRAM could be in principle up to 1000 times faster than the usual Von Neumann architecture.

Finally, we emphasize again that an actual realization of DCRAM is not limited to the employment of solid-state memcapacitive systems considered in this work. Other memcapacitive  systems could serve as even better solutions for practical implementations of DCRAM. We thus hope that our results will be of interest to a wide community of researchers and lead to the next generation of brain-like computing memory.

\section{Acknowledgment}

This work has been partially supported by the Spanish Project TEC2011-14253-E, NSF grants No. DMR-0802830 and ECCS-1202383 and the Center for Magnetic Recording Research at UCSD.

\end{document}